\documentclass[aps,prd,twocolumn,preprintnumbers,floatfix,nofootinbib,showpacs]{revtex4-1}
\usepackage{graphicx}
\usepackage{bm}
\usepackage{times}
\usepackage{amsmath}
\usepackage{slashed}
\usepackage{color}
\usepackage{natbib}
\usepackage{color}
\usepackage{hyperref}
\usepackage[]{appendix}
\usepackage{multirow}
\usepackage{tabularx}
\usepackage{aas_macros}
\usepackage{placeins}
\usepackage[caption=false]{subfig}

\begin{document}

\title{When Dark Matter interacts with Cosmic Rays or Interstellar Matter: A Morphological Study 
}

\author{Eric Carlson}\email{erccarls@ucsc.edu}   
\author{Stefano Profumo}\email{profumo@ucsc.edu}

\affiliation{Department of Physics and Santa Cruz Institute for Particle Physics
University of California, Santa Cruz, CA 95064, USA
}

\date{\today}
\vspace{1cm}

\begin{abstract}
Excess emission over expected diffuse astrophysical backgrounds in the direction of the Galactic center region has been claimed at various wavelengths, from radio to gamma rays. Among particle models advocated to explain such observations, several invoke interactions between dark matter particles and ordinary matter, such as cosmic rays, interstellar gas or free electrons. Depending on the specific interstellar matter particles' species and energy, such models predict distinct morphological features. In this study we make detailed predictions for the morphology of models where the relevant electromagnetic emission is proportional to the product of the dark matter density profile and the density of interstellar matter or cosmic rays. We compare the predicted latitudinal and longitudinal distributions with observations, and provide the associated set of relevant spatial templates.
\end{abstract}

\maketitle

\section{Introduction}
With the advent of large-scale sky surveys at frequencies spanning most of the electromagnetic spectrum, from radio to gamma rays, our theoretical understanding of astrophysical diffuse electromagnetic emission processes is confronting the test of observation in an unprecedented detailed way. Perhaps not surprisingly, at several frequencies diffuse emission models have, at times, fallen short of providing a satisfactory match to observations. Interestingly, in many cases such shortcomings are centered towards the inner regions of the Galaxy, a rich, and relatively poorly understood region. Over the years, some such excesses have found plausible explanations in the realm of ``traditional'' astrophysical processes, or in previously under-estimated emission from populations of astrophysical objects. In some cases, however, it has also been argued that the detected excess might have a ``non-traditional'' origin, possibly connected with new physics.

Perhaps the longest-standing and most widely known such excess is the 511 keV line detected from a broad angular region by INTEGRAL/SPI \cite{511keV}. At larger energy, COMPTEL reported an excess across the energy range between 1-20 MeV \cite{comptel}. Diffuse X-ray emission from the Galactic bulge region, with an approximately thermal spectrum with a very large associated plasma temperature (around 10 keV), has also been reported from Chandra data after point-source subtraction \cite{chandra}. At radio frequencies WMAP revealed excess microwave emission at frequencies between 23 and 61 GHz, an excess known as WMAP haze \cite{Hooper2007}. This radio ``haze'' has also been confirmed with Planck observations \cite{Abergel2011}. Finally, several groups have identified an extended excess of gamma rays from the Galactic center region and from the inner Galaxy, in the few GeV range \cite{Daylan:2014,Calore:2014,Abazajian:2014fta}.

While astrophysical counterparts have been identified that might explain in part or entirely the excesses listed above, several studies have focused on the possible connection of the observed excess emissions with new physics, and specifically with particle dark matter. Dark matter pair-annihilation or decay produces electromagnetic emission both as a result of prompt photon emission from, e.g., neutral pion decay or internal bremsstrahlung or loop-mediated direct annihilation into photons, as well as from secondary mechanisms: the latter mechanism depends on how electrons and positrons, produced in dark matter annihilation or decay, loose energy via synchrotron, inverse Compton, Coulomb scattering and bremsstrahlung \cite{ullioprofumo, Colafrancesco:2005ji}.

As far as the primary emission is concerned, the predicted morphology follows the integral along the line of sight of the dark matter number density (squared) for decay (annihilation, respectively). For the secondary emission, instead, the morphology is complicated by the magnetic field structure and gas and electron densities \cite{Colafrancesco:2006he}.

Alternately, some of the excesses listed above have been associated with slightly less trivial new physics models, where the dark matter electromagnetic emission effectively depends on the environmental cosmic-ray population. Such models include, for example, antiquark nuggets \cite{Lawson2013, zhit}, which would emit via several different mechanisms: free electrons would annihilate with positrons in the nugget's electrosphere producing a 511 keV line; more energetic cosmic ray electrons would penetrate deeper and potentially produce photons in the COMPTEL energy range; cosmic-ray protons penetrating into the quark matter would produce hadronic jets potentially responsible for Bremshstralung emission in the X-ray frequencies relevant for the Chandra excess; for proton cosmic rays penetrating deeper in the nugget, the complete absorption would eventually yield  thermal photons with energies in the WMAP haze range (for a detailed review of all these mechanisms, see Ref.~\cite{zhit}).

 In eXcited dark matter (XDM)~\cite{2007PhRvD..76h3519F} scenarios an excited state exists at energies of a few MeV above the ground state. Such state might be populated by collisions of the dark matter particle with e.g. Galactic cosmic rays, yielding subsequent electron-positron pairs that could explain the 511 keV line signal~\cite{2007PhRvD..76h3519F}.

Other scenarios where electromagnetic emission originates from elastic scattering of dark matter particles off of cosmic rays (or vice versa) were considered in Ref.~\cite{Gorchtein:2010xa,Profumo:2011jt,Profumo:2013jeb}, albeit no specific connection with any of the diffuse excesses was attempted. The relevant cosmic-ray populations are, in this case, high-energy cosmic-ray protons and electrons \cite{Profumo:2013jeb}.

The generic feature of the class of models we focus on here is that the electromagnetic emission is proportional to the integral along the line of sight of the dark matter density times the density of a charged cosmic ray species. In this paper, we consider four classes of such cosmic ray species: (i) low-energy protons; (ii) low-energy free electrons; (iii) intermediate energy (1 GeV) cosmic-ray electrons and protons; and (iv) high-energy (1 TeV) cosmic-ray electrons and protons. We then compare the predicted average longitudinal and latitudinal intensity profiles for the four cases (in some instances we even adopt more than one model for a given case) both with the morphological prediction for dark matter annihilation and with the observed excesse' emission intensity.

The remainder of this study is structured as follows: in sec.~\ref{sec:density} we describe the dark matter and cosmic-ray densities we employ in our analysis; sec.~\ref{sec:profiles} details on the calculation of the resulting emission profiles, while in sec.~\ref{sec:comparison} we compare our findings with the observed excesses' profiles. The final section \ref{sec:disc} presents our discussion and conclusions.

\section{Density Distributions}\label{sec:density}
In this section we describe each of the matter distributions used throughout this study, including the density distribution of dark matter (sec.~\ref{subsec:dm_distributions}), as motivated by the results of current generation N-body simulations; the density of Galactic cosmic-rays for energies between a few hundred MeV and several TeV (sec.~\ref{subsec:Gas_distributions}), as derived through numerical simulations of cosmic-ray propagation; the distributions of interstellar gas (sec.~\ref{subsec:CR_distributions} in two recent and distinct modeling approaches;  and, finally, the most up-to-date model of the Galactic distribution of free electrons (sec.~\ref{subsec:FE_distributions}).  For dark matter, cosmic rays, and gas we also discuss the dominant sources of uncertainty and attempt to bracket the range of state-of-the art models.

\subsection{Dark Matter}
\label{subsec:dm_distributions}

We employ as  our benchmark dark matter profile a Navarro-Frenk-White (NFW)~\cite{NFW:1996} profile with inner slope $\alpha=1$ and scaling radius $r_s=20$~kpc.  In order to assess the uncertainties due to choice of dark matter profiles, we also utilize a steeper (more `cusped') generalized NFW with an inner slope of $\alpha=1.2$; this is in part motivated by recent measurements of the GeV excess~\cite{Daylan:2014,Calore:2014,Abazajian:2014fta}; a steeper inner slope is physically motivated in the context of halo evolution including adiabatic contraction.  On the opposite extreme, we consider a cored Einasto~\cite{NFW:2004} profile with $\alpha_E=0.16$.  We are not concerned with relative normalizations here since we are exclusively interested in the morphological predictions of the models under consideration, not the overall emission intensity.  The functional forms of the two dark matter density profiles we consider here are as follows:

\begin{align}
\rho(r)&=\left(\frac{r_s}{r}\right)^{\alpha} \frac{\rho_0}{(1+r/r_s)^{3-\alpha}},\\[-.7\baselineskip] 
\tag*{NFW} \\
\rho(r)&=\rho_0 \exp{\left( \frac{-2}{\alpha}\left[\left(\frac{r}{r_s}\right)^{-\alpha_E}-1\right]    \right)},\\[-.7\baselineskip] 
\tag*{Einasto}
\end{align}

In Figure~\ref{fig:gas_maps} we show, in the right-most top panel, the integrated line-of-sight dark matter density squared (i.e. the morphology corresponding to an annihilating dark matter candidate) for the case of the NFW profile.  This morphology is of course azimuthally symmetric, and sharply centrally peaked.  In the following two sections we will compare this benchmark scenario to more exotic morphologies arising from dark matter interactions with cosmic-ray protons and electrons, interstellar gas, and free electrons.

\subsection{Interstellar Gas}
\label{subsec:Gas_distributions}
Some models of dark matter -- see e.g. Ref.~\cite{Lawson2013} for a review -- predict interactions with Galactic gas and/or free electrons.  Both of these distributions are strongly peaked toward the Galactic center (GC), and a detailed understanding of the inner few kpc of the Galaxy are needed to formulate solid predictions for the resulting morphology.

The gas density in the Milky Way is typically described by summing contributions from three dynamically distinct components of hydrogen gas in molecular, atomic, and ionized phases. The former two overwhelmingly dominate the gas density near the GC, and are therefore the most important here.

The three-dimensional distribution of gas in the Galaxy can be determined with excellent accuracy by combining surveys of atomic transition lines with a Galactic rotation curve.  For atomic hydrogen, the hyperfine transition at 21 cm provides a direct observable. In the optically thin limit the column density is related to the observed brightness temperature by a single parameter: the hydrogen spin temperature $T_S$. In the case of molecular hydrogen, the lack of a permanent dipole moment requires, instead, use of a tracer gas, the ${\rm CO} (J=1 \to 0)$ transition, which is related to the molecular hydrogen density through a conversion factor $X_{\rm CO}$.  This factor is, in principle, spatially varying.

The deconvolution technique described above relies on a relative velocity between the gas and the solar system.  In the direction of the GC, the gas is co-rotating, implying no kinematic resolution, leading to a distance degeneracy along lines of sight near Galactic longitudes $l\approx0$.  This problem is compounded by the so-called `near-far ambiguity' which corresponds two distances to the same radial velocity in the inner Galaxy.  In order to alleviate such problems, one can incorporate a model into the deconvolution procedure.  For CO ($\rm H_2$ by proxy), we use the gas model from Pohl, Englmaier and Bissantz (PEB), Ref.~\cite{PEB}, which combines the survey of Dame et al~\cite{Dame:2001} with a gas flow model derived from hydrodynamic simulations as well as interpolation of the spiral arms across the line-of-sight toward the GC.  Not only does this provide a significant improvement in the l.o.s. gas distribution toward the Galactic center, but it importantly reconstructs a prominent Galactic bar and the intervening spiral arms. For atomic hydrogen, Nakanishi and Sofue (NS), Ref.~\cite{NS},  assume, instead an analytic model of hydrodynamic equilibrium for the scale height of HI as a function of Galactic radius and uses the Leiden-Argentine-Bonn survey~\cite{LAB} 21 cm survey. This combination provides a high resolution model of $H_2$ and HI which we denote PEB+NS.

One can alternatively attempt to build a three-dimensional model based on observations of individual structures that are prominent in the region. This is the approach of Refs.\cite{Ferriere2001,Ferriere2007} which predicate two primary disk components in the inner 3 kpc of the Galaxy: a dense ``Central Molecular Zone'' approximately centered on the GC and aligned with the Galactic Plane, as well as a holed ``Galactic-Bulge-disk'' which is rotated at $13.5^\circ$ counter-clockwise to the Galactic plane as well as having it's major axis inclined $\approx45^\circ$ away from the line of sight, which is notably much larger than the PEB+NS case, leading to a larger projected extent. Beyond the inner 3 kpc, we use an azimuthally averaged gas profile from Ref.~\cite{NS_H2} to describe the molecular phase, and the HI profile from Ref~\cite{Gordon1976}, corrected to the updated normalizations of Ref.~\cite{Dickey1990}.  We collectively refer to this model as Ferri\`{e}re 2007 (F07). 

For both the PEB+NS and F07 models, we incorporate the contributions of dark gas -- i.e. molecular hydrogen not traced by CO emission, and atomic hydrogen missed due to the assumption of a uniform spin temperature -- by re-normalizing the total `analytic' HI column density along each line of sight to that of the {\tt GALPROP} map \footnote{rbands\_hi12\_v2\_qdeg\_zmax1\_Ts150\_EBV\_mag5\_limit.fits.gz}.  Essentially, this involves a fitting, in addition to HI and H2, a dust template~\cite{SFD:1998} to gamma-ray data. The detailed construction of this template is described in detail in Refs.~\cite{Grenier:2005, Ackermann2012}.

Finally, for both models, we also include a contribution from warm, hot, and very hot phases of ionized hydrogen based on the NE2001~\cite{Cordes2002,cordes2} model for free electrons, described below.  The specific implementation we use is described in Refs.~\cite{Ferriere2001,Ferriere2007}. 

We note that the precise gas distribution in the outer galaxy is of only marginal importance due to the central peak in the dark matter (DM) density profile.  Consequently, we assume both $T_s=150K$ and $X_{\rm CO}=2\times 10^{20}$ $\rm cm^{-2} (K~km/s)^{-1}$ and neglect spatial dependence of either one of these quantities. Although spatial variations are certainly present in both components~\cite{MS:2004, 1978ApJS...36...77D}, we expect the effect of this on the morphology of the Gas$\times$DM profile to be small in the GC region of interest.

\subsection{Cosmic-Rays}
\label{subsec:CR_distributions}

In order to obtain the steady state distribution of cosmic-rays in the Galaxy, we use the numerical code {\tt GALPROP v54.r2504}\footnote{Current versions of {\tt GALPROP} are available at http://galprop.sourceforge.net/}~\cite{galprop1,galprop2}, which encompasses all of the physics relevant for cosmic-ray transport through the galaxy including energy losses, primary source distributions, and re-acceleration.  A detailed description of the physics can be found at the dedicated Web Site\footnote{http://galprop.stanford.edu}. 

Our default diffusion setup consists the model $\rm ^SL^Z4^R20^T150^C5$ in Ref~\cite{Ackermann2012}.  This model features a standard set of diffusion parameters fit to local observations of a variety of primary and secondary species including protons, helium, electrons, positrons, B/C, and $\rm ^{10}Be/^9Be$ \cite{Ackermann2012}. The simulations assume cylindrical geometry with free escape boundary conditions from a diffusion halo of radius 20 kpc and half-height 4 kpc.  Because the dark matter profiles are strongly centrally peaked, most of the reasonable variations related to the diffusion setup and geometry are unimportant to the final cosmic ray (CR) morphology.  This includes the halo height, provided $z_{\rm max}$ is greater than a few kpc and the CR energy of interest is less than a few TeV, at which point CR propagation transitions to the rectilinear regime. Notably, since we are concerned with the morphology and not the spectrum of cosmic rays, parameters describing diffusive re-acceleration and injection spectra are irrelevant. The energy dependence of the diffusion constant will have only a very minimal effect on morphology at the very lowest and highest energies.  For definiteness, we use here a diffusion coefficient $D(\mathcal{R})=5.3\times 10^{28} {\rm cm^2 s^{-1}} (\mathcal{R}/4~{\rm GV})^{1/3}$

Of particular interest to this analysis is the distribution of primary cosmic-ray sources.  In what follows, we show predictions for four tracers of supernova remnants, believed to generate the majority of Galactic cosmic-rays. These are based on the observed surface densities of Galactic pulsars (Yusifov~\cite{Yusifov:2004} and Lorimer~\cite{Lorimer:2006}), OB stars~\cite{Bronfman:2000}, and supernova remnants (SNR~\cite{Case:1998}).  Each of these distributions suffer from substantial uncertainties in the inner 1-2 kpc due to both statistics and systematics surrounding correction for selection effects.  In fact, the latter three distributions are parametrized such that the surface density is zero at the Galactic center, while the Yusifov case is non-zero, leading to a potentially large difference in CR densities in this region.  Simulations of neutron star populations in the Milky Way's gravitational potential also indicate a non-zero central density~\cite{Sartore:2010}, with certain models predicting a strong peak.  Finally, it is difficult to rule out the possibility of a cataclysmic event, or enhanced injection of cosmic-rays at the GC, as may be evidenced by the Fermi Bubbles~\cite{2010ApJ...724.1044S}, and perhaps the Galactic Center $\gamma$-ray Excess itself (see e.g. Refs.~\cite{Carlson:2014, Petrovic:2014}).  Fortunately, only for low kinetic energies -- and only for very high energies for the case of $e^\pm$ -- does this have a significant impact on the CR distributions, which become substantially smoothed after diffusion. In what follows, we do not include these more speculative CR source distributions.

\subsection{Free Electrons}
\label{subsec:FE_distributions}

The density of free electrons $n_e$ in the Milky Way has been mapped to reasonably good precision by fitting complex multicomponent models to a thousands of measurements including primarily pulsar dispersion measures (the line-of-sight integral of the $n_e$), temporal and angular broadening of radio pulses sensitive to variations in $n_e$, and emission measures (the line-of-sight integral of $n_e^2$).  Combining this huge body of information has lead to the development of the so-called NE2001 model~\cite{Cordes2002,cordes2}.

NE2001 contains 5 component classes: (i) smooth components including a thin and thick disc as well as 4 logarithmic spiral arms, (ii) a Galactic center region consisting of a Gaussian component in radius and scale height, (iii) a 4 component local ISM region, (iv) 78 clumped HII regions, and (v) 16 void regions.  The only region with high DM density is the Galactic center, and thus we only include contributions from the dense and thin (140 pc) disk, the lower density thick disk, and the small, but very dense Galactic center components.  We neglect the spiral arms, which do not extend to the inner 3 kpc of the Galaxy as well as clumps and voids which provide only weak and small angular scale ($\theta\ll 1^\circ$) perturbations to the overall density profile in the GC direction.

We also utilize a correction to the thick disk scale-height proposed in Ref.~\cite{Gaensler2008} after recalibrating the NE2001 model while avoiding strong HII regions in the Galactic plane.  This correction roughly doubles the height of the thick disk to $\approx 2$~kpc, which for our purposes leads to a substantial broadening of the expected emission to high latitudes and is therefore less disk-like and more spherically symmetric.  Ref.~\cite{2012MNRAS.427..664S} provides a recent comparison and recalibration of all models of the free electron density. Importantly, they are all based on approximately the same ingredients, with the primary differences in the smooth components being the thick disk scale-height.  The model used here is the thickest, and thus offers the most optimistic scenario for fitting a DM $\times$ free electron signal to an approximately spherical excess.  As we will see below, the primary motivation for modeling such emission is to explain the sharp excess observed at 511 keV by INTEGRAL/SPI and even this optimistic model is too disk-like to explain the signal.

As a final note, the next generation of free electron models are likely to incorporate all-sky surveys of the hydrogen-$\alpha$ emission line, perhaps leading to a significant modification of the model presented here.  Under the assumption that the number density of ionized hydrogen (i.e. free protons) is equal to the number density of free electrons, the H$\alpha$ intensity is directly proportional to the emission measure~\cite{Reynolds1991}.  Unfortunately, such corrections are non-trivial and prone to large errors due to the quadratic dependence of the integrand on $n_e$.

\section{Morphological Profiles}\label{sec:profiles}
\label{sec:Morpholgy}

\begin{figure*}[ht]
\begin{center}

\subfloat{
\includegraphics[width=.9\textwidth]{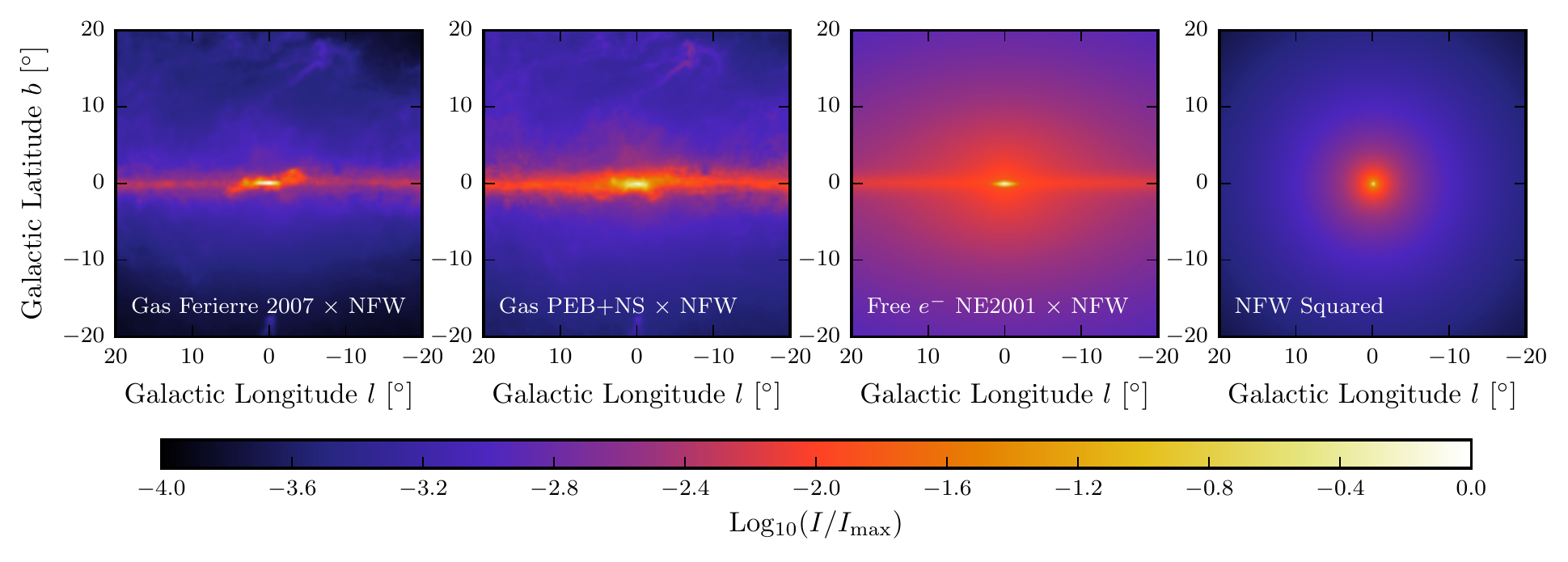}
} \\ \vspace*{-1.4em}
\subfloat{
\includegraphics[width=.9\textwidth]{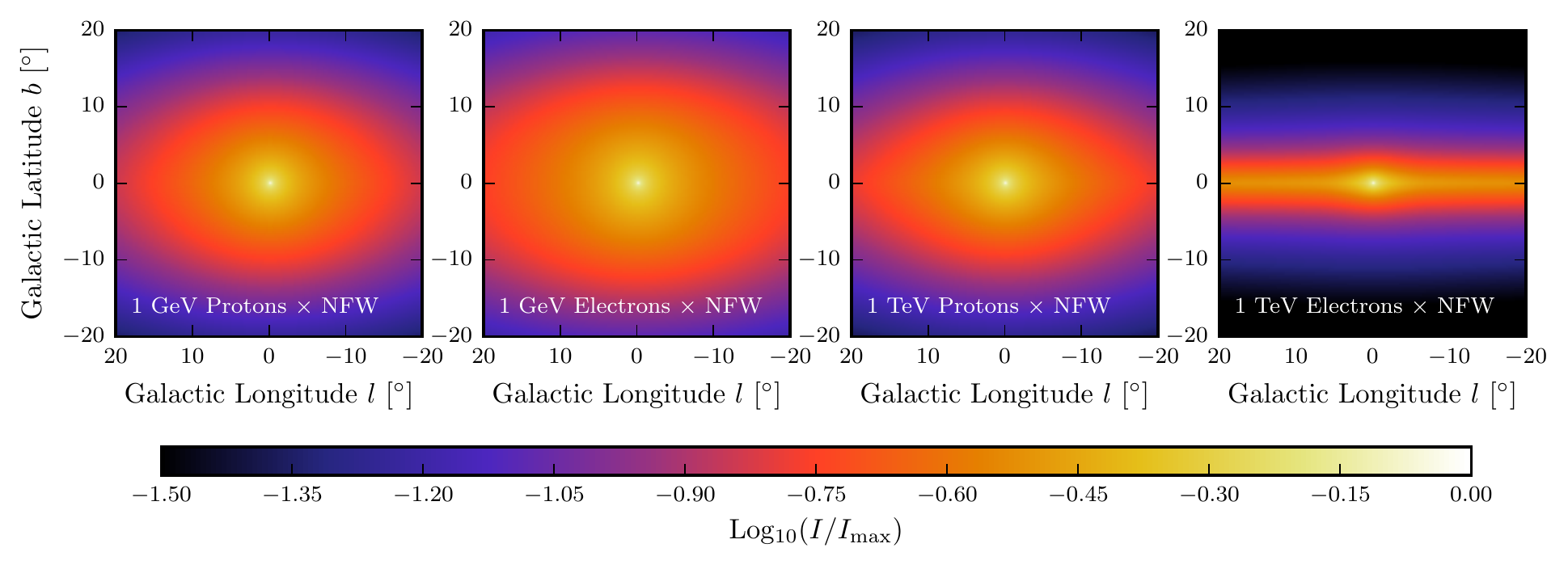}
}

\end{center}

\caption{{\em Top, from left to right:} Projected emission profiles for DM (NFW profile) times the F07 gas model, PEB+NS gas model, free electrons, and, for reference a NFW$^2$ profile (DM annihilation) . \emph{Bottom:} Projected profiles of DM times cosmic-ray protons and electrons for representative energies of 1 GeV (two left-most panels) and 1 TeV (right-most two panels) as calculated with the {\tt GALPROP} package, using the Yusifov~\cite{Yusifov:2004} distribution of primary sources.}
\label{fig:gas_maps}
\end{figure*}

With the matter distributions as specified above, we can now perform the line-of-sight convolution   of dark matter with our gas models (meaning the integral along the line of sigh of dark matter times relevant cosmic ray or gas density). In the top row of Figure~\ref{fig:gas_maps} we show the projected emission profiles for an NFW profile convolved with the F07 and PEB+NS gas models, the NE2001 free electron model, and a standard NFW profile in order to compare against a standard annihilation morphology.  In each case, use of either a contracted NFW or Einasto profile would lead to a slightly brighter and cuspier, or fainter and more cored profile in the innermost regions and each template has been normalized to its maximal value, with no other salient differences.

The F07 gas model possesses a very bright central core due to the highly concentrated CMZ zone surrounding the galactic nucleus (not to be confused with the even denser circum-nuclear ring~\cite{Ferriere:2012} which occupies the innermost $\approx$10 pc and is below the scale probed here).  Emission from the bulge disk can be seen in the diagonally oriented flares on either side. Beyond 5 degrees from the GC, the thin molecular and atomic disks dominate the emission and do not significantly extent to high latitudes making it nearly impossible to obtain a spherical excess.  Similarly, the empirically derived PEB+NS model, shows a bright central disk which more smoothly falls off into the broader Galactic disks.  In addition, the high latitude emission is significantly enhanced with respect to F07. Still, the overall profile lacks azimuthal symmetry and we can conclude that a truly spherical excess is not well fit by DM $\times$ gas profile.

The case of free electrons is more subtle.  Here, we observe a thick disk extending beyond 2 kpc ($b\gtrsim 15^\circ$) which, combined with the DM halo produces a roughly spherical emission profile.  A very bright emission disk can be observed at the Galactic center, though the angular extent is less than  The thin disk, however adds a distinct elongation along the plane making the averaged longitude profile significantly less steep.

In the bottom row of Figure~\ref{fig:gas_maps} we show projected emission templates for benchmark cosmic-ray protons and electrons at 1 GeV and 1 TeV, using a Yusifov profile for the primary source distribution.  For low and high energy protons, as well as low energy electrons, the CR density is relatively uniform over the region of interest, leading approximately to the same profile as would be expected from dark matter \emph{decays}.  For electrons and positrons, inverse-Compton and synchrotron energy loss timescales ($\tau_{\rm ics,sync}\propto E^{-2}$) limit the diffusion radius to $R_{\rm diff} = \sqrt{D(E)\tau_{\rm ics,sync}}\approx 7.5{\rm kpc}/E_{\rm GeV}$ for our choice of parameters, implying that the CR distributions will depend significantly on energy above a few tens of GeV as the cosmic-rays lose energy before propagating farther than a few kpc from their production region (provided the primary source distribution is not uniform).  Supernova remnants are highly concentrated in the plane of the Galaxy, resulting in a full-width-half-max of only a few hundred parsecs near 1 TeV.  Therefore {\em any} dark matter model which predicts emission due to interactions with high energy electrons or positrons will result in a significantly disk-like morphology.

An important caveat should be kept in mind before excluding models based solely on a disk-like component.  Namely that most ``excess'' signals rely on fitting and subtracting off a complicated background model especially at low Galactic latitudes. In some cases, particularly those with gas correlated backgrounds, this fitting procedure can potentially also subtract off a disk component which was actually part of the signal. For example, the background models used to obtain the Galactic Center excess rely on fitting independent gas annuli to gamma-ray data without including any model for excess emission. Thus if the Galactic center excess actually contains a disk-like (gas correlated) component this could be `hiding' in the artificially enhanced background normalization.  One way to alleviate such issues is to fit all components simultaneously rather than relying on residual emission alone.

Finally, we provide skymaps in FITS file format as well as Python scripts at the supplemental materials web page\footnote{\href{http://planck.ucsc.edu/dmcr-morphology}{http://planck.ucsc.edu/dmcr-morphology} } for each combination of halo profile/gas model/free electron model, as well as logarithmically spaced templates in CR energy. Details about the files and model assumptions are provided on the same web page.  We also provide example scripts demonstrating how to specify and integrate an arbitrary three-dimensional distribution against the cosmic-ray, free $e^-$, and gas models which may be found useful in other contexts.

\section{Comparison to Observed Galactic Center Excesses}\label{sec:comparison}
\label{sec:excesses}

\begin{figure*}[ht]\label{fig:511}
\centering
\includegraphics[width=.8\textwidth]{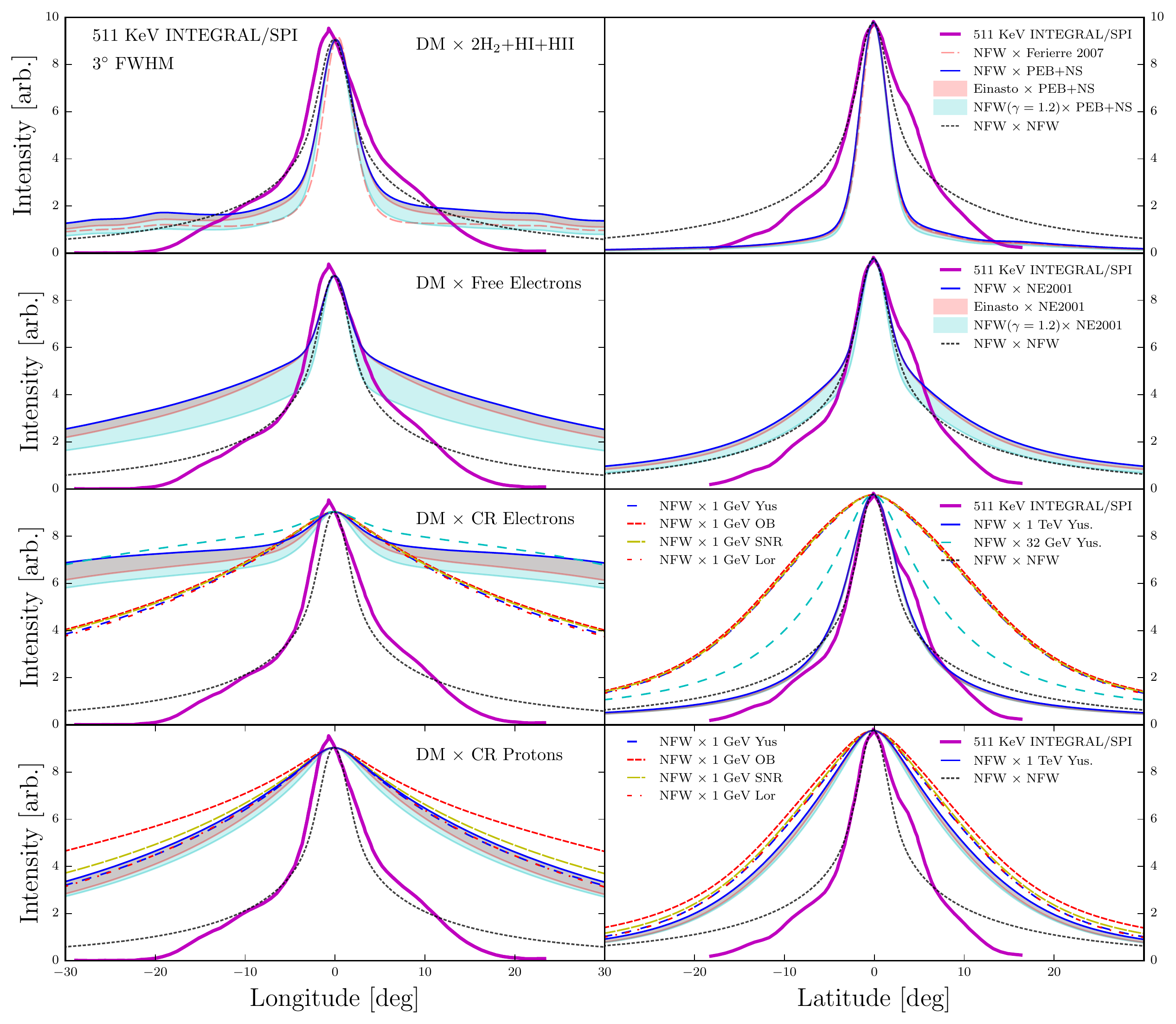}
\caption{Profiles in Galactic longitude (left column) and latitude (right column) of the 511 keV excess detected by INTEGRAL/SPI taken from Ref.~\cite{511keV}.  From the top to bottom row, we show projected profiles for the dark matter density convolved with various distributions (see text) of Galactic gas, free electrons, cosmic-ray electrons, and cosmic ray protons.  The shaded regions show variations due to the choice of dark matter density profile for the benchamark NFW model (solid blue). Each profile has been convolved with a Gaussian with FWHM $3^\circ$ to approximate the INTEGRAL/SPI point spread function, and has been normalized to the same value at $l,b=0$.  The profiles are averaged over $|b|,|l|<30^\circ$, respectively; notice that no error bars are given in Ref.~\cite{511keV}.  }
\label{fig:integral}
\end{figure*}

We now proceed to comparing the morphology resulting from the models under consideration here to three residual signals observed in the Galactic center region. 

The first signal of interest is the 511 keV line. First detected in 1972\cite{1972ApJ...172L...1J}, the 511 keV excess has since been the subject of lively scientific debate.  The most reliable measurement~\cite{511keV} is provided by the SPI instrument on-board the ESA's INTEGRAL satellite which indicates a large excess population of low-energy positrons in the Galactic center region. Proposed explanations range from cosmic-ray interactions with the interstellar medium to exotic astrophysical objects, to dark matter\footnote{For an extensive review of this signal and its possible origins we refer the reader to Ref.~\cite{Prantzos2010}.}.  However, any plausible explanation must not only explain the spectrum, but also the basic morphological properties of the observed excess which are well well fit by two components: (i) a spherical-bulge consisting of two radial Gaussians (projected FWHMs of $3^\circ$ and $11^\circ$) and a $\approx$30\% fainter thick disk (FWHM $7^\circ$)~\cite{Weidenspointner2008}.

In Figure \ref{fig:511} we compare longitude and latitude profiles~\cite{511keV} of the 511 keV excess in the Galactic center region against each of the four morphological profiles. To account for the finite angular resolution of the instrument, we have applied a Gaussian point spread function with a FWHM of $3^\circ$.  The longitude and latitude profiles have been averaged over $|b|,|l|<30^\circ$, respectively, in order to match the same angular averages as in the reported excess of Ref.~\cite{511keV}. Each distribution has been normalized independently in both latitude and longitude at $l,b=0$ which is suitable for observations with a large point spread function.  In each panel we also show the prediction for a squared NFW profile, for comparison to the standard case of annihilating dark matter.  To represent variations on the dark matter halo, we show the canonical convolution with an NFW profile in dark blue, with shaded cyan (red) regions representing the contracted NFW (Einasto) profiles.

\begin{figure*}
\centering
\includegraphics[width=.8\textwidth]{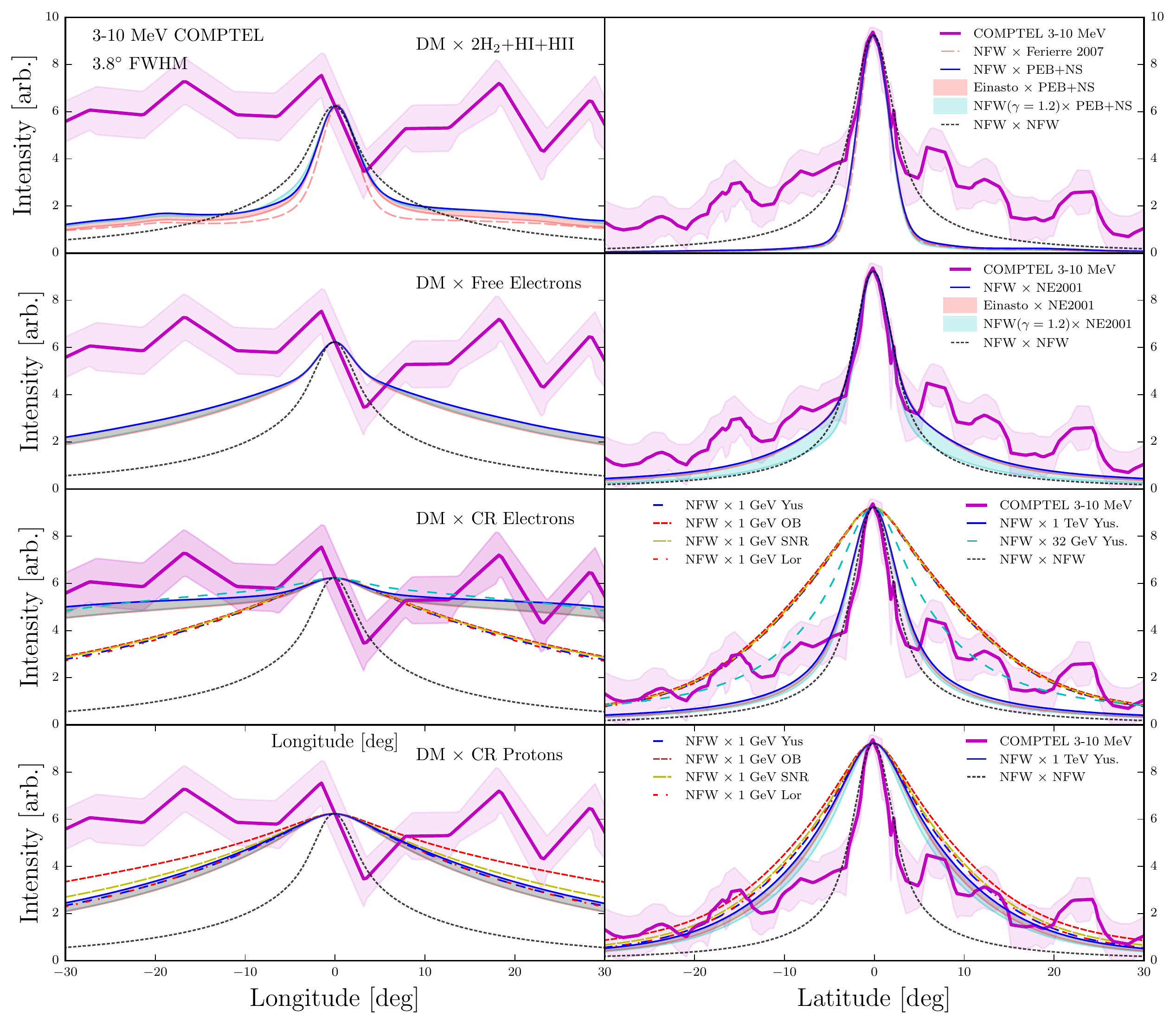}
\caption{Same as Figure~\ref{fig:integral}, but for the 3-10 MeV COMPTEL excess of Ref.~\cite{comptel}, after subtracting the average `base' emission. Predicted profiles have been convoluted with a PSF of FWHM=$3.8^\circ$, and have been averaged over $|b|<5^\circ$ and $1^\circ<|l|<30^\circ$, respectively.}
\label{fig:comptel}
\end{figure*}

In the top row of Fig.~ \ref{fig:511} we show the DM$\times$Gas for our PEB+NS and F07 models.  As observed in the previous section, the latitude profile is extremely centrally peaked and is much to steep to match the observed excess. Likewise, the longitude profile is also very centrally peaked due to the very bright CMZ region surrounding the GC. This is particularly true in the F07 model, where off-center emission is orders of magnitude lower beyond $5^\circ$. Variations in the halo profile are relatively small and unable to reconcile a compatible signal. 

In the second row, we show the free-electron case. Here the halo model has a much larger impact due to the extended distribution of free $e^-$.  Because 511 keV corresponds to the electron/positron mass, DM$\times$Free $e^\pm$ is perhaps the primary case of interest for this signal, as DM models producing an abundance of low energy positrons will form para-positronium bound states which subsequently decay into two 511 MeV $\gamma$-rays.  Another possibility arises in models of Quark `(anti)-Nugget' Dark Matter~\cite{Lawson2013}, in which heavy `nuggets' of quark matter form as a result of a new high-density color-superconducting QCD phase.  Such models can explain both baryogenisis and provide a connection to the $\approx5:1$ relative abundance of dark matter and baryons. In these models quark matter can be surrounded by a layer of positrons which can annihilate through collisions with free electrons.  As noted in the Introduction, these models also predict a variety of other multi-wavelength signatures which are tested below and act as an interesting test case throughout for confirming or ruling out models based on the detailed morphology of DM-matter interactions. Our results indicate that for any angular scale larger than a few degrees, the predicted morphology for DM$\times$Free electrons does not match the observed signal, with significant over-predicted intensity around 10 degrees in longitude.

In the third row, we show the emission profile for cosmic-ray electrons at 1, 32, and $10^3$ GeV and protons at 1 GeV and 1 TeV, along with the variations introduced by CR source distributions. In eXcited DM (XDM)~\cite{2007PhRvD..76h3519F} scenarios it has also been proposed that a DM particle is collisionally excited through some inelastic interaction with, e.g. cosmic-rays or gas, into a higher energy state. If the difference in energy levels is $\delta E=2m_e$, such a state could decay through emission of a non-relativistic $e^+e^-$ pair, where the $e^+$ subsequently annihilates.  Our results indicate that for either protons or electrons the predicted morphology is qualitatively incompatible in both latitude and longitude with the observed 511 keV signal profile. The morphology predicted for pair-annihilating DM models provides a much better fit to the observed morphology.

We now consider observations from the Compton Telescope (COMPTEL), which detects a significant and broad spectrum excess from 1-20 MeV which is well over the expected inverse-Compton and bremsstrahlung backgrounds. Limited statistics and systematics surrounding the deconvolution techniques lead to a somewhat noisy morphological profile which is difficult to definitively call diffuse, as opposed to some combination of point sources (e.g. hard X-rays from supernovae).  Models exist which postulate a sharp steepening in the cosmic-ray electron spectrum below 200 MeV exist which do not violate synchrotron constraints~\cite{2000ApJ...537..763S} and can potentially also link the hard X-ray excess from INTEGRAL to the low energy $\gamma$-ray regime (EGRET).  However, such models lack a straightforward physical interpretation and are remain {\em ad hoc}, leaving open the door for new physics explanations.  A variety of astrophysical models are reviewed in Ref.~\cite{2000ApJ...537..763S}.

Figure \ref{fig:comptel} shows our results for the COMPTEL 3-10 MeV excess, where the average ``base'' emission has been subtracted off \cite{comptel}. The predicted emission profiles are again normalized to the $l,b=0$ value, and have been averaged over $|b|<5^\circ$ and $1^\circ<|l|<30^\circ$, respectively and smoother out with a Gaussian PSF of FWHM=$3.8^\circ$ as warranted for the COMPTEL observations. The four rows and two columns adopt the same conventions as in Fig.~\ref{fig:511}.

In the context of the anti-quark nugget model, the COMPTEL emission might be associated with non-relativistic cosmic ray electrons  penetrating deeper in the nugget's electrosphere. As discussed in Ref.~\cite{Forbes:2009wg} the relevant electron velocities to explain the COMPTEL excess spectrum range between $v/c=0.001$ and 0.01. Therefore, the relevant population corresponds to the free electrons case (second row), although it is interesting to also compare with the cosmic-ray electron case (third row). While the CR electron longitudinal profile is marginally compatible with the flat COMPTEL profile, the free electron case is significantly under-abundant. The latitudinal profile, instead, shows a remarkably close match for the free electron case to the observed profile, but only in the innermost few degrees, while falling short at larger latitudes. The cosmic-ray electron  (or proton) latitudinal profiles are instead significantly different from the observed excess. Finally, convolution with gas density profiles also provides a decent match to observations as a function of latitude in the innermost few degrees.

In summary, models invoking a convolution of free electrons and dark matter to explain the COMPTEL excess, such as antiquark nuggets, provide an adequate morphological shape for the innermost regions at low latitude, but fall short at larger latitudes and provide a relatively poor match to the longitudinal profile.

\begin{figure}\label{fig:lat}
\centering
\includegraphics[width=.45\textwidth]{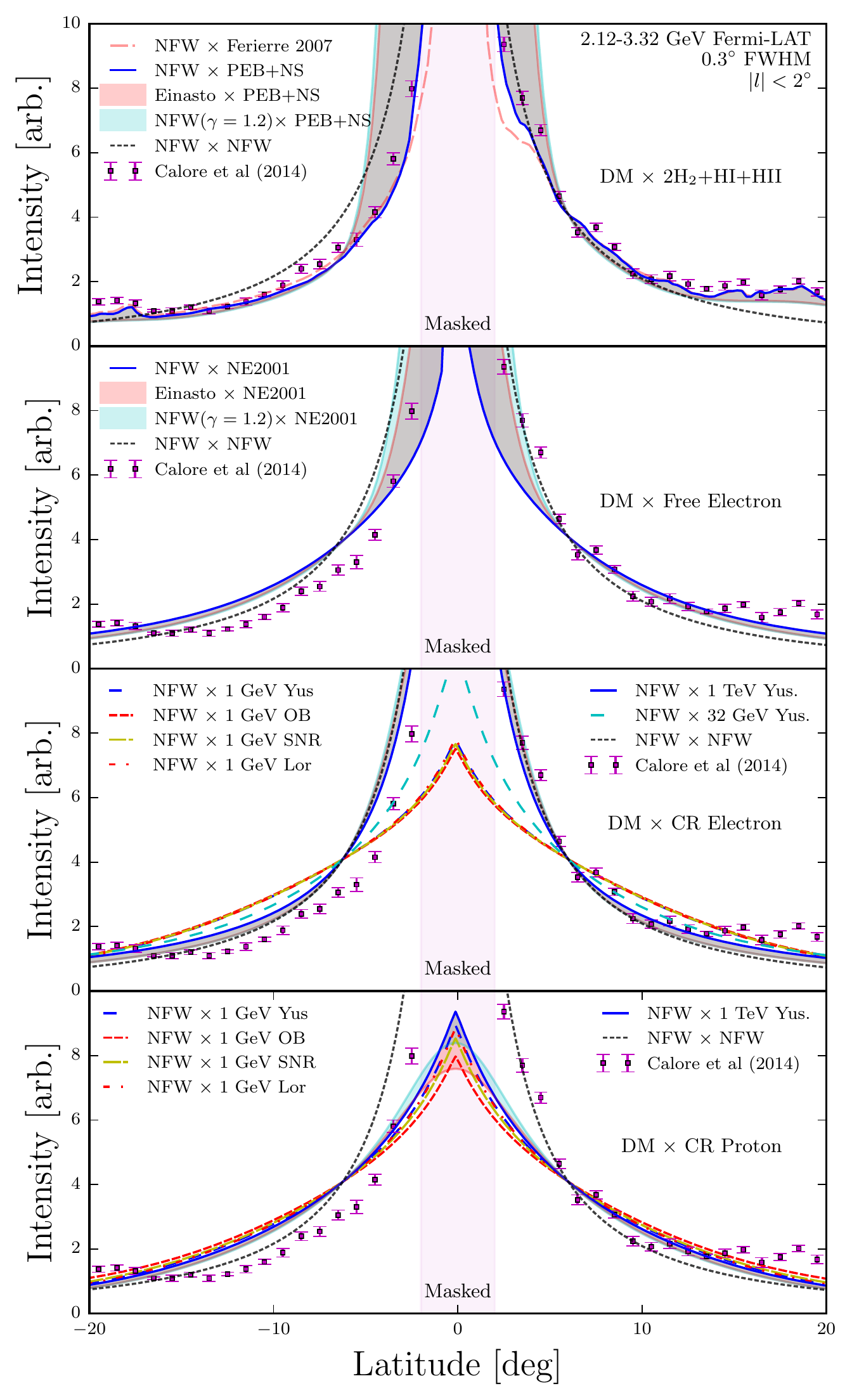}
\caption{Same as Figure~\ref{fig:integral}, but for the $\sim$GeV Fermi-LAT `excess' from Ref.~\cite{Calore:2014}. Predicted profiles use a PSF of FWHM=$0.3^\circ$, are averaged over $|l|<2^\circ$, and are normalized at $b=+6^\circ$.}
\label{fig:GCE}
\end{figure}

Finally, we come to the observed Fermi-LAT GeV excess.  Detected initially in 2009~\cite{Goodenough:2009gk}, a recent, detailed re-evaulation~\cite{Daylan:2014} has led to renewed interest and subsequent detection by several other groups~\cite{Calore2014, Abazajian:2014fta}.  Based on current state-of-the art diffuse $\gamma$-ray background models, the excess is approximately spherical with a radial profile consistent with a contracted NFW.  The spectrum is also well fit by a $\approx 50$ GeV weakly interacting massive particle annihilating to $b$-quarks with a cross-section near the thermal relic value, making the signal a prime candidate for the potential detection of dark matter.  Thus far, proposed astrophysical explanations in terms of unresolved pulsars~\cite{gordon_macias:2013} have failed to explain the hard low-energy spectum, luminosity, and spatial extent~\cite{Cholis:2014lta}, which has now been detected beyond $15^\circ$~\cite{Calore2014} (see however, the recent Ref.~\cite{O'Leary:2015gfa} which proposes a new population of very young pulsars).  A viable interpretation involving cosmic-ray outbursts from the Galactic center region~\cite{Carlson:2014,Petrovic:2014} remains an open question, though for hadronic models the significant gas correlated emission from neutral pion decay imposes substantial constraints.

In Figure \ref{fig:lat} we exhibit our results for the latitudinal profile from the Galactic center taken from Ref.~\cite{Calore:2014}. Here, we smooth the predicted profiles with a PSF of FWHM=$0.3^\circ$, and we average, as in Ref.~\cite{Calore:2014} over $|l|<2^\circ$. Unlike the previous cases we normalize the predicted emission profiles at $b=+6^\circ$.

Our key findings are that models predicating a convolution of DM with interstellar gas or free electrons produce a remarkably good fit to the latitudinal distribution of the observed profile (top two rows). Models, instead, involving DM$\times$CR convolutions do not provide a good match, over predicting signals at large latitudes and falling short in the inner regions --- in other words producing unacceptably shallow density profiles. This rules out for example interpretation based on the models discussed in Ref.~\cite{Profumo:2011jt}.  Although longitudinal distributions are currently unavailable, the observed excess has been found to be highly spherical making such that any profile with a significant disk component is unlikely to be viable barring substantial changes to the diffuse background model.

\section{Discussion and Conclusions}\label{sec:disc}
We carried out a model-independent study of scenarios where diffuse electromagnetic emission originates from interactions of dark matter particles with the interstellar gas, free electrons or Galactic cosmic rays. We assumed that the relevant morphology depends on the integrated line-of-sight product of the dark matter particle density times the relevant gas/cosmic-ray density. The key motivation to consider such models is that a variety of large-scale diffuse excesses in the general direction of the Galactic center have been identified, at wavelengths ranging from radio to gamma rays, and that a variety of particle physics scenarios have been proposed that rely on dark matter interacting with Galactic gas or cosmic rays.

We considered a variety of state-of-the-art gas density models, and well-motivated cosmic ray models, utilizing several different assumed injection source distribution profiles. We produced the relevant latitudinal and longitudinal profiles, and we are making our results available publicly on the web. Finally, we compared our predictions with the emission profiles of a few selected diffuse excesses.

Generally, we find that the morphology of the 511 keV line is matched much more closely by dark matter pair-annihilation than by any of the dark matter times gas/cosmic ray models we investigated; the COMPTEL excess is reproduced by models of dark matter times gas or free electrons in the innermost regions, at low Galactic latitude, but not at large latitudes or in longitude. Finally, models advocating dark matter/cosmic-ray convolution fail at reproducing the Fermi-LAT GeV excess, while dark matter times gas or free electrons provides a relatively good fit to the latitudinal distribution of the signal.

%\FloatBarrier

\section{Acknowledgments}
EC is supported by a NASA Graduate Research Fellowship under NASA NESSF Grant No. NNX13AO63H. SP is partly supported by the US Department of Energy, Contract DE-FG02- 04ER41286.

\bibliography{morpho}

\begin{thebibliography}{59}
\expandafter\ifx\csname natexlab\endcsname\relax\def\natexlab#1{#1}\fi
\expandafter\ifx\csname bibnamefont\endcsname\relax
  \def\bibnamefont#1{#1}\fi
\expandafter\ifx\csname bibfnamefont\endcsname\relax
  \def\bibfnamefont#1{#1}\fi
\expandafter\ifx\csname citenamefont\endcsname\relax
  \def\citenamefont#1{#1}\fi
\expandafter\ifx\csname url\endcsname\relax
  \def\url#1{\texttt{#1}}\fi
\expandafter\ifx\csname urlprefix\endcsname\relax\def\urlprefix{URL }\fi
\providecommand{\bibinfo}[2]{#2}
\providecommand{\eprint}[2][]{\url{#2}}

\bibitem[{\citenamefont{Kn\"{o}dlseder
  et~al.}(2005)\citenamefont{Kn\"{o}dlseder, Jean, Lonjou, Weidenspointner,
  Guessoum, Gillard, Skinner, von Ballmoos, Vedrenne, Roques et~al.}}]{511keV}
\bibinfo{author}{\bibfnamefont{J.}~\bibnamefont{Kn\"{o}dlseder}},
  \bibinfo{author}{\bibfnamefont{P.}~\bibnamefont{Jean}},
  \bibinfo{author}{\bibfnamefont{V.}~\bibnamefont{Lonjou}},
  \bibinfo{author}{\bibfnamefont{G.}~\bibnamefont{Weidenspointner}},
  \bibinfo{author}{\bibfnamefont{N.}~\bibnamefont{Guessoum}},
  \bibinfo{author}{\bibfnamefont{W.}~\bibnamefont{Gillard}},
  \bibinfo{author}{\bibfnamefont{G.}~\bibnamefont{Skinner}},
  \bibinfo{author}{\bibfnamefont{P.}~\bibnamefont{von Ballmoos}},
  \bibinfo{author}{\bibfnamefont{G.}~\bibnamefont{Vedrenne}},
  \bibinfo{author}{\bibfnamefont{J.-P.} \bibnamefont{Roques}},
  \bibnamefont{et~al.}, \bibinfo{journal}{Astronomy and Astrophysics}
  \textbf{\bibinfo{volume}{441}}, \bibinfo{pages}{513} (\bibinfo{year}{2005}),
  ISSN \bibinfo{issn}{0004-6361}, \eprint{0506026v1},
  \urlprefix\url{http://arxiv.org/pdf/astro-ph/0506026v1.pdf
  http://www.edpsciences.org/10.1051/0004-6361:20042063}.

\bibitem[{\citenamefont{{Strong} et~al.}(1999)\citenamefont{{Strong},
  {Bloemen}, {Diehl}, {Hermsen}, and {Sch{\"o}nfelder}}}]{comptel}
\bibinfo{author}{\bibfnamefont{A.~W.} \bibnamefont{{Strong}}},
  \bibinfo{author}{\bibfnamefont{H.}~\bibnamefont{{Bloemen}}},
  \bibinfo{author}{\bibfnamefont{R.}~\bibnamefont{{Diehl}}},
  \bibinfo{author}{\bibfnamefont{W.}~\bibnamefont{{Hermsen}}},
  \bibnamefont{and}
  \bibinfo{author}{\bibfnamefont{V.}~\bibnamefont{{Sch{\"o}nfelder}}},
  \bibinfo{journal}{Astrophysical Letters and Communications}
  \textbf{\bibinfo{volume}{39}}, \bibinfo{pages}{209} (\bibinfo{year}{1999}),
  \eprint{astro-ph/9811211}.

\bibitem[{\citenamefont{Muno et~al.}(2004)\citenamefont{Muno, Baganoff, Bautz,
  Feigelson, Garmire et~al.}}]{chandra}
\bibinfo{author}{\bibfnamefont{M.~P.} \bibnamefont{Muno}},
  \bibinfo{author}{\bibfnamefont{F.}~\bibnamefont{Baganoff}},
  \bibinfo{author}{\bibfnamefont{M.}~\bibnamefont{Bautz}},
  \bibinfo{author}{\bibfnamefont{E.}~\bibnamefont{Feigelson}},
  \bibinfo{author}{\bibfnamefont{G.}~\bibnamefont{Garmire}},
  \bibnamefont{et~al.}, \bibinfo{journal}{Astrophys.J.}
  \textbf{\bibinfo{volume}{613}}, \bibinfo{pages}{326} (\bibinfo{year}{2004}),
  \eprint{astro-ph/0402087}.

\bibitem[{\citenamefont{Hooper et~al.}(2007)\citenamefont{Hooper, Finkbeiner,
  and Dobler}}]{Hooper2007}
\bibinfo{author}{\bibfnamefont{D.}~\bibnamefont{Hooper}},
  \bibinfo{author}{\bibfnamefont{D.~P.} \bibnamefont{Finkbeiner}},
  \bibnamefont{and} \bibinfo{author}{\bibfnamefont{G.}~\bibnamefont{Dobler}},
  p.~\bibinfo{pages}{4} (\bibinfo{year}{2007}), \eprint{0705.3655},
  \urlprefix\url{http://arxiv.org/pdf/0705.3655v1.pdf
  http://arxiv.org/abs/0705.3655}.

\bibitem[{\citenamefont{Abergel et~al.}(2011)\citenamefont{Abergel, Ade,
  Aghanim, Arnaud, Ashdown, Aumont, Baccigalupi, and Balbi}}]{Abergel2011}
\bibinfo{author}{\bibfnamefont{P.~C.~A.} \bibnamefont{Abergel}},
  \bibinfo{author}{\bibfnamefont{P.~A.~R.} \bibnamefont{Ade}},
  \bibinfo{author}{\bibfnamefont{N.}~\bibnamefont{Aghanim}},
  \bibinfo{author}{\bibfnamefont{M.}~\bibnamefont{Arnaud}},
  \bibinfo{author}{\bibfnamefont{M.}~\bibnamefont{Ashdown}},
  \bibinfo{author}{\bibfnamefont{J.}~\bibnamefont{Aumont}},
  \bibinfo{author}{\bibfnamefont{C.}~\bibnamefont{Baccigalupi}},
  \bibnamefont{and} \bibinfo{author}{\bibfnamefont{A.}~\bibnamefont{Balbi}},
  \textbf{\bibinfo{volume}{24}}, \bibinfo{pages}{1} (\bibinfo{year}{2011}),
  \urlprefix\url{http://www.aanda.org/articles/aa/pdf/2011/12/aa16485-11.pdf}.

\bibitem[{\citenamefont{{Daylan} et~al.}(2014)\citenamefont{{Daylan},
  {Finkbeiner}, {Hooper}, {Linden}, {Portillo}, {Rodd}, and
  {Slatyer}}}]{Daylan:2014}
\bibinfo{author}{\bibfnamefont{T.}~\bibnamefont{{Daylan}}},
  \bibinfo{author}{\bibfnamefont{D.~P.} \bibnamefont{{Finkbeiner}}},
  \bibinfo{author}{\bibfnamefont{D.}~\bibnamefont{{Hooper}}},
  \bibinfo{author}{\bibfnamefont{T.}~\bibnamefont{{Linden}}},
  \bibinfo{author}{\bibfnamefont{S.~K.~N.} \bibnamefont{{Portillo}}},
  \bibinfo{author}{\bibfnamefont{N.~L.} \bibnamefont{{Rodd}}},
  \bibnamefont{and} \bibinfo{author}{\bibfnamefont{T.~R.}
  \bibnamefont{{Slatyer}}}, \bibinfo{journal}{ArXiv e-prints}
  (\bibinfo{year}{2014}), \eprint{1402.6703}.

\bibitem[{\citenamefont{{Calore} et~al.}(2014)\citenamefont{{Calore}, {Cholis},
  and {Weniger}}}]{Calore:2014}
\bibinfo{author}{\bibfnamefont{F.}~\bibnamefont{{Calore}}},
  \bibinfo{author}{\bibfnamefont{I.}~\bibnamefont{{Cholis}}}, \bibnamefont{and}
  \bibinfo{author}{\bibfnamefont{C.}~\bibnamefont{{Weniger}}},
  \bibinfo{journal}{ArXiv e-prints}  (\bibinfo{year}{2014}),
  \eprint{1409.0042}.

\bibitem[{\citenamefont{Abazajian et~al.}(2014)\citenamefont{Abazajian, Canac,
  Horiuchi, and Kaplinghat}}]{Abazajian:2014fta}
\bibinfo{author}{\bibfnamefont{K.~N.} \bibnamefont{Abazajian}},
  \bibinfo{author}{\bibfnamefont{N.}~\bibnamefont{Canac}},
  \bibinfo{author}{\bibfnamefont{S.}~\bibnamefont{Horiuchi}}, \bibnamefont{and}
  \bibinfo{author}{\bibfnamefont{M.}~\bibnamefont{Kaplinghat}},
  \bibinfo{journal}{Physical Review D} \textbf{\bibinfo{volume}{90}},
  \bibinfo{pages}{023526} (\bibinfo{year}{2014}), ISSN
  \bibinfo{issn}{1550-7998}, \eprint{1402.4090},
  \urlprefix\url{http://link.aps.org/doi/10.1103/PhysRevD.90.023526}.

\bibitem[{\citenamefont{Profumo and Ullio}(2010)}]{ullioprofumo}
\bibinfo{author}{\bibfnamefont{S.}~\bibnamefont{Profumo}} \bibnamefont{and}
  \bibinfo{author}{\bibfnamefont{P.}~\bibnamefont{Ullio}}
  (\bibinfo{year}{2010}), \eprint{1001.4086}.

\bibitem[{\citenamefont{Colafrancesco et~al.}(2006)\citenamefont{Colafrancesco,
  Profumo, and Ullio}}]{Colafrancesco:2005ji}
\bibinfo{author}{\bibfnamefont{S.}~\bibnamefont{Colafrancesco}},
  \bibinfo{author}{\bibfnamefont{S.}~\bibnamefont{Profumo}}, \bibnamefont{and}
  \bibinfo{author}{\bibfnamefont{P.}~\bibnamefont{Ullio}},
  \bibinfo{journal}{Astron.Astrophys.} \textbf{\bibinfo{volume}{455}},
  \bibinfo{pages}{21} (\bibinfo{year}{2006}), \eprint{astro-ph/0507575}.

\bibitem[{\citenamefont{Colafrancesco et~al.}(2007)\citenamefont{Colafrancesco,
  Profumo, and Ullio}}]{Colafrancesco:2006he}
\bibinfo{author}{\bibfnamefont{S.}~\bibnamefont{Colafrancesco}},
  \bibinfo{author}{\bibfnamefont{S.}~\bibnamefont{Profumo}}, \bibnamefont{and}
  \bibinfo{author}{\bibfnamefont{P.}~\bibnamefont{Ullio}},
  \bibinfo{journal}{Phys.Rev.} \textbf{\bibinfo{volume}{D75}},
  \bibinfo{pages}{023513} (\bibinfo{year}{2007}), \eprint{astro-ph/0607073}.

\bibitem[{\citenamefont{Lawson and
  Zhitnitsky}(2013{\natexlab{a}})}]{Lawson2013}
\bibinfo{author}{\bibfnamefont{K.}~\bibnamefont{Lawson}} \bibnamefont{and}
  \bibinfo{author}{\bibfnamefont{A.~R.} \bibnamefont{Zhitnitsky}}, pp.
  \bibinfo{pages}{1--7} (\bibinfo{year}{2013}{\natexlab{a}}),
  \eprint{1305.6318}, \urlprefix\url{http://arxiv.org/pdf/1305.6318v2.pdf
  http://arxiv.org/abs/1305.6318}.

\bibitem[{\citenamefont{Lawson and Zhitnitsky}(2013{\natexlab{b}})}]{zhit}
\bibinfo{author}{\bibfnamefont{K.}~\bibnamefont{Lawson}} \bibnamefont{and}
  \bibinfo{author}{\bibfnamefont{A.~R.} \bibnamefont{Zhitnitsky}}
  (\bibinfo{year}{2013}{\natexlab{b}}), \eprint{1305.6318}.

\bibitem[{\citenamefont{{Finkbeiner} and {Weiner}}(2007)}]{2007PhRvD..76h3519F}
\bibinfo{author}{\bibfnamefont{D.~P.} \bibnamefont{{Finkbeiner}}}
  \bibnamefont{and} \bibinfo{author}{\bibfnamefont{N.}~\bibnamefont{{Weiner}}},
  \bibinfo{journal}{\prd} \textbf{\bibinfo{volume}{76}}, \bibinfo{eid}{083519}
  (\bibinfo{year}{2007}), \eprint{astro-ph/0702587}.

\bibitem[{\citenamefont{Gorchtein et~al.}(2010)\citenamefont{Gorchtein,
  Profumo, and Ubaldi}}]{Gorchtein:2010xa}
\bibinfo{author}{\bibfnamefont{M.}~\bibnamefont{Gorchtein}},
  \bibinfo{author}{\bibfnamefont{S.}~\bibnamefont{Profumo}}, \bibnamefont{and}
  \bibinfo{author}{\bibfnamefont{L.}~\bibnamefont{Ubaldi}},
  \bibinfo{journal}{Phys.Rev.} \textbf{\bibinfo{volume}{D82}},
  \bibinfo{pages}{083514} (\bibinfo{year}{2010}), \eprint{1008.2230}.

\bibitem[{\citenamefont{Profumo and Ubaldi}(2011)}]{Profumo:2011jt}
\bibinfo{author}{\bibfnamefont{S.}~\bibnamefont{Profumo}} \bibnamefont{and}
  \bibinfo{author}{\bibfnamefont{L.}~\bibnamefont{Ubaldi}},
  \bibinfo{journal}{JCAP} \textbf{\bibinfo{volume}{1108}}, \bibinfo{pages}{020}
  (\bibinfo{year}{2011}), \eprint{1106.4568}.

\bibitem[{\citenamefont{Profumo et~al.}(2013)\citenamefont{Profumo, Ubaldi, and
  Gorchtein}}]{Profumo:2013jeb}
\bibinfo{author}{\bibfnamefont{S.}~\bibnamefont{Profumo}},
  \bibinfo{author}{\bibfnamefont{L.}~\bibnamefont{Ubaldi}}, \bibnamefont{and}
  \bibinfo{author}{\bibfnamefont{M.}~\bibnamefont{Gorchtein}},
  \bibinfo{journal}{JCAP} \textbf{\bibinfo{volume}{1304}}, \bibinfo{pages}{012}
  (\bibinfo{year}{2013}), \eprint{1302.1915}.

\bibitem[{\citenamefont{{Navarro} et~al.}(1996)\citenamefont{{Navarro},
  {Frenk}, and {White}}}]{NFW:1996}
\bibinfo{author}{\bibfnamefont{J.~F.} \bibnamefont{{Navarro}}},
  \bibinfo{author}{\bibfnamefont{C.~S.} \bibnamefont{{Frenk}}},
  \bibnamefont{and} \bibinfo{author}{\bibfnamefont{S.~D.~M.}
  \bibnamefont{{White}}}, \bibinfo{journal}{\apj}
  \textbf{\bibinfo{volume}{462}}, \bibinfo{pages}{563} (\bibinfo{year}{1996}),
  \eprint{astro-ph/9508025}.

\bibitem[{\citenamefont{{Navarro} et~al.}(2004)\citenamefont{{Navarro},
  {Hayashi}, {Power}, {Jenkins}, {Frenk}, {White}, {Springel}, {Stadel}, and
  {Quinn}}}]{NFW:2004}
\bibinfo{author}{\bibfnamefont{J.~F.} \bibnamefont{{Navarro}}},
  \bibinfo{author}{\bibfnamefont{E.}~\bibnamefont{{Hayashi}}},
  \bibinfo{author}{\bibfnamefont{C.}~\bibnamefont{{Power}}},
  \bibinfo{author}{\bibfnamefont{A.~R.} \bibnamefont{{Jenkins}}},
  \bibinfo{author}{\bibfnamefont{C.~S.} \bibnamefont{{Frenk}}},
  \bibinfo{author}{\bibfnamefont{S.~D.~M.} \bibnamefont{{White}}},
  \bibinfo{author}{\bibfnamefont{V.}~\bibnamefont{{Springel}}},
  \bibinfo{author}{\bibfnamefont{J.}~\bibnamefont{{Stadel}}}, \bibnamefont{and}
  \bibinfo{author}{\bibfnamefont{T.~R.} \bibnamefont{{Quinn}}},
  \bibinfo{journal}{\mnras} \textbf{\bibinfo{volume}{349}},
  \bibinfo{pages}{1039} (\bibinfo{year}{2004}), \eprint{astro-ph/0311231}.

\bibitem[{\citenamefont{{Pohl} et~al.}(2008)\citenamefont{{Pohl}, {Englmaier},
  and {Bissantz}}}]{PEB}
\bibinfo{author}{\bibfnamefont{M.}~\bibnamefont{{Pohl}}},
  \bibinfo{author}{\bibfnamefont{P.}~\bibnamefont{{Englmaier}}},
  \bibnamefont{and}
  \bibinfo{author}{\bibfnamefont{N.}~\bibnamefont{{Bissantz}}},
  \bibinfo{journal}{\apj} \textbf{\bibinfo{volume}{677}}, \bibinfo{pages}{283}
  (\bibinfo{year}{2008}), \eprint{0712.4264}.

\bibitem[{\citenamefont{{Dame} et~al.}(2001)\citenamefont{{Dame}, {Hartmann},
  and {Thaddeus}}}]{Dame:2001}
\bibinfo{author}{\bibfnamefont{T.~M.} \bibnamefont{{Dame}}},
  \bibinfo{author}{\bibfnamefont{D.}~\bibnamefont{{Hartmann}}},
  \bibnamefont{and}
  \bibinfo{author}{\bibfnamefont{P.}~\bibnamefont{{Thaddeus}}},
  \bibinfo{journal}{\apj} \textbf{\bibinfo{volume}{547}}, \bibinfo{pages}{792}
  (\bibinfo{year}{2001}), \eprint{astro-ph/0009217}.

\bibitem[{\citenamefont{{Nakanishi} and {Sofue}}(2003)}]{NS}
\bibinfo{author}{\bibfnamefont{H.}~\bibnamefont{{Nakanishi}}} \bibnamefont{and}
  \bibinfo{author}{\bibfnamefont{Y.}~\bibnamefont{{Sofue}}},
  \bibinfo{journal}{\pasj} \textbf{\bibinfo{volume}{55}}, \bibinfo{pages}{191}
  (\bibinfo{year}{2003}), \eprint{astro-ph/0304338}.

\bibitem[{\citenamefont{{Kalberla} et~al.}(2005)\citenamefont{{Kalberla},
  {Burton}, {Hartmann}, {Arnal}, {Bajaja}, {Morras}, and {P{\"o}ppel}}}]{LAB}
\bibinfo{author}{\bibfnamefont{P.~M.~W.} \bibnamefont{{Kalberla}}},
  \bibinfo{author}{\bibfnamefont{W.~B.} \bibnamefont{{Burton}}},
  \bibinfo{author}{\bibfnamefont{D.}~\bibnamefont{{Hartmann}}},
  \bibinfo{author}{\bibfnamefont{E.~M.} \bibnamefont{{Arnal}}},
  \bibinfo{author}{\bibfnamefont{E.}~\bibnamefont{{Bajaja}}},
  \bibinfo{author}{\bibfnamefont{R.}~\bibnamefont{{Morras}}}, \bibnamefont{and}
  \bibinfo{author}{\bibfnamefont{W.~G.~L.} \bibnamefont{{P{\"o}ppel}}},
  \bibinfo{journal}{\aap} \textbf{\bibinfo{volume}{440}}, \bibinfo{pages}{775}
  (\bibinfo{year}{2005}), \eprint{astro-ph/0504140}.

\bibitem[{\citenamefont{Ferriere}(2001)}]{Ferriere2001}
\bibinfo{author}{\bibfnamefont{K.~M.} \bibnamefont{Ferriere}},
  p.~\bibinfo{pages}{39} (\bibinfo{year}{2001}), \eprint{0106359},
  \urlprefix\url{http://arxiv.org/pdf/astro-ph/0106359.pdf
  http://arxiv.org/abs/astro-ph/0106359}.

\bibitem[{\citenamefont{Ferri\`{e}re et~al.}(2007)\citenamefont{Ferri\`{e}re,
  Gillard, and Jean}}]{Ferriere2007}
\bibinfo{author}{\bibfnamefont{K.}~\bibnamefont{Ferri\`{e}re}},
  \bibinfo{author}{\bibfnamefont{W.}~\bibnamefont{Gillard}}, \bibnamefont{and}
  \bibinfo{author}{\bibfnamefont{P.}~\bibnamefont{Jean}},
  \bibinfo{journal}{Astronomy and Astrophysics} \textbf{\bibinfo{volume}{467}},
  \bibinfo{pages}{611} (\bibinfo{year}{2007}), ISSN \bibinfo{issn}{0004-6361},
  \eprint{0702532}, \urlprefix\url{http://arxiv.org/pdf/astro-ph/0702532v1.pdf
  http://arxiv.org/abs/astro-ph/0702532
  http://www.aanda.org/10.1051/0004-6361:20066992}.

\bibitem[{\citenamefont{{Nakanishi} and {Sofue}}(2006)}]{NS_H2}
\bibinfo{author}{\bibfnamefont{H.}~\bibnamefont{{Nakanishi}}} \bibnamefont{and}
  \bibinfo{author}{\bibfnamefont{Y.}~\bibnamefont{{Sofue}}},
  \bibinfo{journal}{\pasj} \textbf{\bibinfo{volume}{58}}, \bibinfo{pages}{847}
  (\bibinfo{year}{2006}), \eprint{astro-ph/0610769}.

\bibitem[{\citenamefont{Gordon and Burton}(1976)}]{Gordon1976}
\bibinfo{author}{\bibfnamefont{M.~A.} \bibnamefont{Gordon}} \bibnamefont{and}
  \bibinfo{author}{\bibfnamefont{W.~B.} \bibnamefont{Burton}},
  \bibinfo{journal}{The Astrophysical Journal} \textbf{\bibinfo{volume}{208}},
  \bibinfo{pages}{346} (\bibinfo{year}{1976}), ISSN \bibinfo{issn}{0004-637X},
  \urlprefix\url{http://adsabs.harvard.edu/abs/1976ApJ...208..346G
  http://adsabs.harvard.edu/doi/10.1086/154613}.

\bibitem[{\citenamefont{Dickey and Lockman}(1990)}]{Dickey1990}
\bibinfo{author}{\bibfnamefont{J.~M.} \bibnamefont{Dickey}} \bibnamefont{and}
  \bibinfo{author}{\bibfnamefont{F.~J.} \bibnamefont{Lockman}},
  \bibinfo{journal}{Annual Review of Astronomy and Astrophysics}
  \textbf{\bibinfo{volume}{28}}, \bibinfo{pages}{215} (\bibinfo{year}{1990}),
  ISSN \bibinfo{issn}{0066-4146},
  \urlprefix\url{http://www.annualreviews.org/doi/pdf/10.1146/annurev.aa.28.090190.001243
  http://www.annualreviews.org/doi/abs/10.1146/annurev.aa.28.090190.001243}.

\bibitem[{\citenamefont{{Schlegel} et~al.}(1998)\citenamefont{{Schlegel},
  {Finkbeiner}, and {Davis}}}]{SFD:1998}
\bibinfo{author}{\bibfnamefont{D.}~\bibnamefont{{Schlegel}}},
  \bibinfo{author}{\bibfnamefont{D.}~\bibnamefont{{Finkbeiner}}},
  \bibnamefont{and} \bibinfo{author}{\bibfnamefont{M.}~\bibnamefont{{Davis}}},
  in \emph{\bibinfo{booktitle}{Wide Field Surveys in Cosmology}}, edited by
  \bibinfo{editor}{\bibfnamefont{S.}~\bibnamefont{{Colombi}}},
  \bibinfo{editor}{\bibfnamefont{Y.}~\bibnamefont{{Mellier}}},
  \bibnamefont{and} \bibinfo{editor}{\bibfnamefont{B.}~\bibnamefont{{Raban}}}
  (\bibinfo{year}{1998}), p. \bibinfo{pages}{297}, \eprint{astro-ph/9809230}.

\bibitem[{\citenamefont{{Grenier} et~al.}(2005)\citenamefont{{Grenier},
  {Casandjian}, and {Terrier}}}]{Grenier:2005}
\bibinfo{author}{\bibfnamefont{I.~A.} \bibnamefont{{Grenier}}},
  \bibinfo{author}{\bibfnamefont{J.-M.} \bibnamefont{{Casandjian}}},
  \bibnamefont{and}
  \bibinfo{author}{\bibfnamefont{R.}~\bibnamefont{{Terrier}}},
  \bibinfo{journal}{Science} \textbf{\bibinfo{volume}{307}},
  \bibinfo{pages}{1292} (\bibinfo{year}{2005}).

\bibitem[{\citenamefont{Ackermann et~al.}(2012)\citenamefont{Ackermann, Ajello,
  Atwood, Baldini, Ballet, Barbiellini, Bastieri, Bechtol, Bellazzini, Berenji
  et~al.}}]{Ackermann2012}
\bibinfo{author}{\bibfnamefont{M.}~\bibnamefont{Ackermann}},
  \bibinfo{author}{\bibfnamefont{M.}~\bibnamefont{Ajello}},
  \bibinfo{author}{\bibfnamefont{W.~B.} \bibnamefont{Atwood}},
  \bibinfo{author}{\bibfnamefont{L.}~\bibnamefont{Baldini}},
  \bibinfo{author}{\bibfnamefont{J.}~\bibnamefont{Ballet}},
  \bibinfo{author}{\bibfnamefont{G.}~\bibnamefont{Barbiellini}},
  \bibinfo{author}{\bibfnamefont{D.}~\bibnamefont{Bastieri}},
  \bibinfo{author}{\bibfnamefont{K.}~\bibnamefont{Bechtol}},
  \bibinfo{author}{\bibfnamefont{R.}~\bibnamefont{Bellazzini}},
  \bibinfo{author}{\bibfnamefont{B.}~\bibnamefont{Berenji}},
  \bibnamefont{et~al.}, \bibinfo{journal}{The Astrophysical Journal}
  \textbf{\bibinfo{volume}{750}}, \bibinfo{pages}{3} (\bibinfo{year}{2012}),
  ISSN \bibinfo{issn}{0004-637X},
  \urlprefix\url{http://stacks.iop.org/0004-637X/750/i=1/a=3?key=crossref.df6dd3ed0cf5a605dffbcb75777753db}.

\bibitem[{\citenamefont{Cordes and Lazio}(2002)}]{Cordes2002}
\bibinfo{author}{\bibfnamefont{J.}~\bibnamefont{Cordes}} \bibnamefont{and}
  \bibinfo{author}{\bibfnamefont{T.}~\bibnamefont{Lazio}},
  \bibinfo{journal}{arXiv preprint astro-ph/0207156} p.~\bibinfo{pages}{21}
  (\bibinfo{year}{2002}), \eprint{0207156},
  \urlprefix\url{http://arxiv.org/pdf/astro-ph/0207156v3.pdf
  http://arxiv.org/abs/astro-ph/0207156 http://arxiv.org/abs/astroph/0207156}.

\bibitem[{\citenamefont{Cordes and Lazio}(2003)}]{cordes2}
\bibinfo{author}{\bibfnamefont{J.~M.} \bibnamefont{Cordes}} \bibnamefont{and}
  \bibinfo{author}{\bibfnamefont{T.~J.~W.} \bibnamefont{Lazio}},
  p.~\bibinfo{pages}{41} (\bibinfo{year}{2003}), \eprint{astro-ph/0301598},
  \urlprefix\url{http://arxiv.org/pdf/astro-ph/0301598v1.pdf
  http://arxiv.org/abs/astro-ph/0301598}.

\bibitem[{\citenamefont{{Strong} et~al.}(2004)\citenamefont{{Strong},
  {Moskalenko}, {Reimer}, {Digel}, and {Diehl}}}]{MS:2004}
\bibinfo{author}{\bibfnamefont{A.~W.} \bibnamefont{{Strong}}},
  \bibinfo{author}{\bibfnamefont{I.~V.} \bibnamefont{{Moskalenko}}},
  \bibinfo{author}{\bibfnamefont{O.}~\bibnamefont{{Reimer}}},
  \bibinfo{author}{\bibfnamefont{S.}~\bibnamefont{{Digel}}}, \bibnamefont{and}
  \bibinfo{author}{\bibfnamefont{R.}~\bibnamefont{{Diehl}}},
  \bibinfo{journal}{\aap} \textbf{\bibinfo{volume}{422}}, \bibinfo{pages}{L47}
  (\bibinfo{year}{2004}), \eprint{astro-ph/0405275}.

\bibitem[{\citenamefont{{Dickey} et~al.}(1978)\citenamefont{{Dickey},
  {Terzian}, and {Salpeter}}}]{1978ApJS...36...77D}
\bibinfo{author}{\bibfnamefont{J.~M.} \bibnamefont{{Dickey}}},
  \bibinfo{author}{\bibfnamefont{Y.}~\bibnamefont{{Terzian}}},
  \bibnamefont{and} \bibinfo{author}{\bibfnamefont{E.~E.}
  \bibnamefont{{Salpeter}}}, \bibinfo{journal}{\apjs}
  \textbf{\bibinfo{volume}{36}}, \bibinfo{pages}{77} (\bibinfo{year}{1978}).

\bibitem[{\citenamefont{{Strong}
  et~al.}(2000{\natexlab{a}})\citenamefont{{Strong}, {Moskalenko}, and
  {Reimer}}}]{galprop1}
\bibinfo{author}{\bibfnamefont{A.~W.} \bibnamefont{{Strong}}},
  \bibinfo{author}{\bibfnamefont{I.~V.} \bibnamefont{{Moskalenko}}},
  \bibnamefont{and} \bibinfo{author}{\bibfnamefont{O.}~\bibnamefont{{Reimer}}},
  \bibinfo{journal}{\apj} \textbf{\bibinfo{volume}{537}}, \bibinfo{pages}{763}
  (\bibinfo{year}{2000}{\natexlab{a}}), \eprint{astro-ph/9811296}.

\bibitem[{\citenamefont{{Strong} and {Moskalenko}}(1998)}]{galprop2}
\bibinfo{author}{\bibfnamefont{A.~W.} \bibnamefont{{Strong}}} \bibnamefont{and}
  \bibinfo{author}{\bibfnamefont{I.~V.} \bibnamefont{{Moskalenko}}},
  \bibinfo{journal}{\apj} \textbf{\bibinfo{volume}{509}}, \bibinfo{pages}{212}
  (\bibinfo{year}{1998}), \eprint{astro-ph/9807150}.

\bibitem[{\citenamefont{{Yusifov} and {K{\"u}{\c
  c}{\"u}k}}(2004)}]{Yusifov:2004}
\bibinfo{author}{\bibfnamefont{I.}~\bibnamefont{{Yusifov}}} \bibnamefont{and}
  \bibinfo{author}{\bibfnamefont{I.}~\bibnamefont{{K{\"u}{\c c}{\"u}k}}}, in
  \emph{\bibinfo{booktitle}{The Magnetized Interstellar Medium}}, edited by
  \bibinfo{editor}{\bibfnamefont{B.}~\bibnamefont{{Uyaniker}}},
  \bibinfo{editor}{\bibfnamefont{W.}~\bibnamefont{{Reich}}}, \bibnamefont{and}
  \bibinfo{editor}{\bibfnamefont{R.}~\bibnamefont{{Wielebinski}}}
  (\bibinfo{year}{2004}), pp. \bibinfo{pages}{159--164},
  \eprint{astro-ph/0405495}.

\bibitem[{\citenamefont{{Lorimer} et~al.}(2006)\citenamefont{{Lorimer},
  {Faulkner}, {Lyne}, {Manchester}, {Kramer}, {McLaughlin}, {Hobbs},
  {Possenti}, {Stairs}, {Camilo} et~al.}}]{Lorimer:2006}
\bibinfo{author}{\bibfnamefont{D.~R.} \bibnamefont{{Lorimer}}},
  \bibinfo{author}{\bibfnamefont{A.~J.} \bibnamefont{{Faulkner}}},
  \bibinfo{author}{\bibfnamefont{A.~G.} \bibnamefont{{Lyne}}},
  \bibinfo{author}{\bibfnamefont{R.~N.} \bibnamefont{{Manchester}}},
  \bibinfo{author}{\bibfnamefont{M.}~\bibnamefont{{Kramer}}},
  \bibinfo{author}{\bibfnamefont{M.~A.} \bibnamefont{{McLaughlin}}},
  \bibinfo{author}{\bibfnamefont{G.}~\bibnamefont{{Hobbs}}},
  \bibinfo{author}{\bibfnamefont{A.}~\bibnamefont{{Possenti}}},
  \bibinfo{author}{\bibfnamefont{I.~H.} \bibnamefont{{Stairs}}},
  \bibinfo{author}{\bibfnamefont{F.}~\bibnamefont{{Camilo}}},
  \bibnamefont{et~al.}, \bibinfo{journal}{\mnras}
  \textbf{\bibinfo{volume}{372}}, \bibinfo{pages}{777} (\bibinfo{year}{2006}),
  \eprint{astro-ph/0607640}.

\bibitem[{\citenamefont{{Bronfman} et~al.}(2000)\citenamefont{{Bronfman},
  {Casassus}, {May}, and {Nyman}}}]{Bronfman:2000}
\bibinfo{author}{\bibfnamefont{L.}~\bibnamefont{{Bronfman}}},
  \bibinfo{author}{\bibfnamefont{S.}~\bibnamefont{{Casassus}}},
  \bibinfo{author}{\bibfnamefont{J.}~\bibnamefont{{May}}}, \bibnamefont{and}
  \bibinfo{author}{\bibfnamefont{L.-{\AA}.} \bibnamefont{{Nyman}}},
  \bibinfo{journal}{\aap} \textbf{\bibinfo{volume}{358}}, \bibinfo{pages}{521}
  (\bibinfo{year}{2000}), \eprint{astro-ph/0006104}.

\bibitem[{\citenamefont{{Case} and {Bhattacharya}}(1998)}]{Case:1998}
\bibinfo{author}{\bibfnamefont{G.~L.} \bibnamefont{{Case}}} \bibnamefont{and}
  \bibinfo{author}{\bibfnamefont{D.}~\bibnamefont{{Bhattacharya}}},
  \bibinfo{journal}{\apj} \textbf{\bibinfo{volume}{504}}, \bibinfo{pages}{761}
  (\bibinfo{year}{1998}), \eprint{astro-ph/9807162}.

\bibitem[{\citenamefont{{Sartore} et~al.}(2010)\citenamefont{{Sartore},
  {Ripamonti}, {Treves}, and {Turolla}}}]{Sartore:2010}
\bibinfo{author}{\bibfnamefont{N.}~\bibnamefont{{Sartore}}},
  \bibinfo{author}{\bibfnamefont{E.}~\bibnamefont{{Ripamonti}}},
  \bibinfo{author}{\bibfnamefont{A.}~\bibnamefont{{Treves}}}, \bibnamefont{and}
  \bibinfo{author}{\bibfnamefont{R.}~\bibnamefont{{Turolla}}},
  \bibinfo{journal}{\aap} \textbf{\bibinfo{volume}{510}}, \bibinfo{eid}{A23}
  (\bibinfo{year}{2010}), \eprint{0908.3182}.

\bibitem[{\citenamefont{{Su} et~al.}(2010)\citenamefont{{Su}, {Slatyer}, and
  {Finkbeiner}}}]{2010ApJ...724.1044S}
\bibinfo{author}{\bibfnamefont{M.}~\bibnamefont{{Su}}},
  \bibinfo{author}{\bibfnamefont{T.~R.} \bibnamefont{{Slatyer}}},
  \bibnamefont{and} \bibinfo{author}{\bibfnamefont{D.~P.}
  \bibnamefont{{Finkbeiner}}}, \bibinfo{journal}{\apj}
  \textbf{\bibinfo{volume}{724}}, \bibinfo{pages}{1044} (\bibinfo{year}{2010}),
  \eprint{1005.5480}.

\bibitem[{\citenamefont{{Carlson} and {Profumo}}(2014)}]{Carlson:2014}
\bibinfo{author}{\bibfnamefont{E.}~\bibnamefont{{Carlson}}} \bibnamefont{and}
  \bibinfo{author}{\bibfnamefont{S.}~\bibnamefont{{Profumo}}},
  \bibinfo{journal}{\prd} \textbf{\bibinfo{volume}{90}}, \bibinfo{eid}{023015}
  (\bibinfo{year}{2014}), \eprint{1405.7685}.

\bibitem[{\citenamefont{{Petrovi{\'c}}
  et~al.}(2014)\citenamefont{{Petrovi{\'c}}, {Dario Serpico}, and {Zaharija{\v
  s}}}}]{Petrovic:2014}
\bibinfo{author}{\bibfnamefont{J.}~\bibnamefont{{Petrovi{\'c}}}},
  \bibinfo{author}{\bibfnamefont{P.}~\bibnamefont{{Dario Serpico}}},
  \bibnamefont{and}
  \bibinfo{author}{\bibfnamefont{G.}~\bibnamefont{{Zaharija{\v s}}}},
  \bibinfo{journal}{\jcap} \textbf{\bibinfo{volume}{10}}, \bibinfo{eid}{052}
  (\bibinfo{year}{2014}), \eprint{1405.7928}.

\bibitem[{\citenamefont{Gaensler}(2008)}]{Gaensler2008}
\bibinfo{author}{\bibfnamefont{B.}~\bibnamefont{Gaensler}},
  \bibinfo{journal}{Publications of the \ldots} pp. \bibinfo{pages}{1--20}
  (\bibinfo{year}{2008}), \eprint{arXiv:0808.2550v1},
  \urlprefix\url{http://arxiv.org/pdf/0808.2550v1.pdf
  http://journals.cambridge.org/abstract\_S1323358000004720}.

\bibitem[{\citenamefont{{Schnitzeler}}(2012)}]{2012MNRAS.427..664S}
\bibinfo{author}{\bibfnamefont{D.~H.~F.~M.} \bibnamefont{{Schnitzeler}}},
  \bibinfo{journal}{\mnras} \textbf{\bibinfo{volume}{427}},
  \bibinfo{pages}{664} (\bibinfo{year}{2012}), \eprint{1208.3045}.

\bibitem[{\citenamefont{Reynolds}(1991)}]{Reynolds1991}
\bibinfo{author}{\bibfnamefont{R.~J.} \bibnamefont{Reynolds}},
  \bibinfo{journal}{The Astrophysical Journal} \textbf{\bibinfo{volume}{372}},
  \bibinfo{pages}{L17} (\bibinfo{year}{1991}), ISSN \bibinfo{issn}{0004-637X},
  \urlprefix\url{http://articles.adsabs.harvard.edu/cgi-bin/nph-iarticle\_query?1991ApJ...372L..17R\&amp;data\_type=PDF\_HIGH\&amp;whole\_paper=YES\&amp;type=PRINTER\&amp;filetype=.pdf
  http://adsabs.harvard.edu/doi/10.1086/186013}.

\bibitem[{\citenamefont{{Ferri{\`e}re}}(2012)}]{Ferriere:2012}
\bibinfo{author}{\bibfnamefont{K.}~\bibnamefont{{Ferri{\`e}re}}},
  \bibinfo{journal}{\aap} \textbf{\bibinfo{volume}{540}}, \bibinfo{eid}{A50}
  (\bibinfo{year}{2012}), \eprint{1201.6031}.

\bibitem[{\citenamefont{{Johnson} et~al.}(1972)\citenamefont{{Johnson},
  {Harnden}, and {Haymes}}}]{1972ApJ...172L...1J}
\bibinfo{author}{\bibfnamefont{W.~N.} \bibnamefont{{Johnson}},
  \bibfnamefont{III}}, \bibinfo{author}{\bibfnamefont{F.~R.}
  \bibnamefont{{Harnden}}, \bibfnamefont{Jr.}}, \bibnamefont{and}
  \bibinfo{author}{\bibfnamefont{R.~C.} \bibnamefont{{Haymes}}},
  \bibinfo{journal}{\apjl} \textbf{\bibinfo{volume}{172}}, \bibinfo{pages}{L1}
  (\bibinfo{year}{1972}).

\bibitem[{\citenamefont{Prantzos et~al.}(2010)\citenamefont{Prantzos, Boehm,
  Bykov, Diehl, Ferriere, Guessoum, Jean, Knoedlseder, Marcowith, Moskalenko
  et~al.}}]{Prantzos2010}
\bibinfo{author}{\bibfnamefont{N.}~\bibnamefont{Prantzos}},
  \bibinfo{author}{\bibfnamefont{C.}~\bibnamefont{Boehm}},
  \bibinfo{author}{\bibfnamefont{A.~M.} \bibnamefont{Bykov}},
  \bibinfo{author}{\bibfnamefont{R.}~\bibnamefont{Diehl}},
  \bibinfo{author}{\bibfnamefont{K.}~\bibnamefont{Ferriere}},
  \bibinfo{author}{\bibfnamefont{N.}~\bibnamefont{Guessoum}},
  \bibinfo{author}{\bibfnamefont{P.}~\bibnamefont{Jean}},
  \bibinfo{author}{\bibfnamefont{J.}~\bibnamefont{Knoedlseder}},
  \bibinfo{author}{\bibfnamefont{A.}~\bibnamefont{Marcowith}},
  \bibinfo{author}{\bibfnamefont{I.~V.} \bibnamefont{Moskalenko}},
  \bibnamefont{et~al.}, p.~\bibinfo{pages}{62} (\bibinfo{year}{2010}),
  \eprint{1009.4620}, \urlprefix\url{http://arxiv.org/pdf/1009.4620v1.pdf
  http://arxiv.org/abs/1009.4620}.

\bibitem[{\citenamefont{Weidenspointner
  et~al.}(2008)\citenamefont{Weidenspointner, Skinner, Jean, Kn\"{o}dlseder,
  von Ballmoos, Bignami, Diehl, Strong, Cordier, Schanne
  et~al.}}]{Weidenspointner2008}
\bibinfo{author}{\bibfnamefont{G.}~\bibnamefont{Weidenspointner}},
  \bibinfo{author}{\bibfnamefont{G.}~\bibnamefont{Skinner}},
  \bibinfo{author}{\bibfnamefont{P.}~\bibnamefont{Jean}},
  \bibinfo{author}{\bibfnamefont{J.}~\bibnamefont{Kn\"{o}dlseder}},
  \bibinfo{author}{\bibfnamefont{P.}~\bibnamefont{von Ballmoos}},
  \bibinfo{author}{\bibfnamefont{G.}~\bibnamefont{Bignami}},
  \bibinfo{author}{\bibfnamefont{R.}~\bibnamefont{Diehl}},
  \bibinfo{author}{\bibfnamefont{A.~W.} \bibnamefont{Strong}},
  \bibinfo{author}{\bibfnamefont{B.}~\bibnamefont{Cordier}},
  \bibinfo{author}{\bibfnamefont{S.}~\bibnamefont{Schanne}},
  \bibnamefont{et~al.}, \bibinfo{journal}{Nature}
  \textbf{\bibinfo{volume}{451}}, \bibinfo{pages}{159} (\bibinfo{year}{2008}),
  ISSN \bibinfo{issn}{1476-4687},
  \urlprefix\url{http://www.ncbi.nlm.nih.gov/pubmed/18185581}.

\bibitem[{\citenamefont{{Strong}
  et~al.}(2000{\natexlab{b}})\citenamefont{{Strong}, {Moskalenko}, and
  {Reimer}}}]{2000ApJ...537..763S}
\bibinfo{author}{\bibfnamefont{A.~W.} \bibnamefont{{Strong}}},
  \bibinfo{author}{\bibfnamefont{I.~V.} \bibnamefont{{Moskalenko}}},
  \bibnamefont{and} \bibinfo{author}{\bibfnamefont{O.}~\bibnamefont{{Reimer}}},
  \bibinfo{journal}{\apj} \textbf{\bibinfo{volume}{537}}, \bibinfo{pages}{763}
  (\bibinfo{year}{2000}{\natexlab{b}}), \eprint{astro-ph/9811296}.

\bibitem[{\citenamefont{Forbes et~al.}(2010)\citenamefont{Forbes, Lawson, and
  Zhitnitsky}}]{Forbes:2009wg}
\bibinfo{author}{\bibfnamefont{M.~M.} \bibnamefont{Forbes}},
  \bibinfo{author}{\bibfnamefont{K.}~\bibnamefont{Lawson}}, \bibnamefont{and}
  \bibinfo{author}{\bibfnamefont{A.~R.} \bibnamefont{Zhitnitsky}},
  \bibinfo{journal}{Phys.Rev.} \textbf{\bibinfo{volume}{D82}},
  \bibinfo{pages}{083510} (\bibinfo{year}{2010}), \eprint{0910.4541}.

\bibitem[{\citenamefont{Goodenough and Hooper}(2009)}]{Goodenough:2009gk}
\bibinfo{author}{\bibfnamefont{L.}~\bibnamefont{Goodenough}} \bibnamefont{and}
  \bibinfo{author}{\bibfnamefont{D.}~\bibnamefont{Hooper}}
  (\bibinfo{year}{2009}), \eprint{0910.2998}.

\bibitem[{\citenamefont{Calore et~al.}(2014)\citenamefont{Calore, Cholis, and
  Weniger}}]{Calore2014}
\bibinfo{author}{\bibfnamefont{F.}~\bibnamefont{Calore}},
  \bibinfo{author}{\bibfnamefont{I.}~\bibnamefont{Cholis}}, \bibnamefont{and}
  \bibinfo{author}{\bibfnamefont{C.}~\bibnamefont{Weniger}},
  p.~\bibinfo{pages}{65} (\bibinfo{year}{2014}), \eprint{1409.0042},
  \urlprefix\url{http://arxiv.org/pdf/1409.0042v1.pdf
  http://arxiv.org/abs/1409.0042}.

\bibitem[{\citenamefont{Gordon and Mac\'{\i}as}(2013)}]{gordon_macias:2013}
\bibinfo{author}{\bibfnamefont{C.}~\bibnamefont{Gordon}} \bibnamefont{and}
  \bibinfo{author}{\bibfnamefont{O.}~\bibnamefont{Mac\'{\i}as}},
  \bibinfo{journal}{$\backslash$prd} \textbf{\bibinfo{volume}{88}},
  \bibinfo{pages}{83521} (\bibinfo{year}{2013}), \eprint{1306.5725}.

\bibitem[{\citenamefont{Cholis et~al.}(2014)\citenamefont{Cholis, Hooper, and
  Linden}}]{Cholis:2014lta}
\bibinfo{author}{\bibfnamefont{I.}~\bibnamefont{Cholis}},
  \bibinfo{author}{\bibfnamefont{D.}~\bibnamefont{Hooper}}, \bibnamefont{and}
  \bibinfo{author}{\bibfnamefont{T.}~\bibnamefont{Linden}}
  (\bibinfo{year}{2014}), \eprint{1407.5625}.

\bibitem[{\citenamefont{O'Leary et~al.}(2015)\citenamefont{O'Leary, Kistler,
  Kerr, and Dexter}}]{O'Leary:2015gfa}
\bibinfo{author}{\bibfnamefont{R.~M.} \bibnamefont{O'Leary}},
  \bibinfo{author}{\bibfnamefont{M.~D.} \bibnamefont{Kistler}},
  \bibinfo{author}{\bibfnamefont{M.}~\bibnamefont{Kerr}}, \bibnamefont{and}
  \bibinfo{author}{\bibfnamefont{J.}~\bibnamefont{Dexter}}
  (\bibinfo{year}{2015}), \eprint{1504.02477}.

\end{thebibliography}
\end{document}